\documentclass[prl,aps,singlecolumn,superscriptaddress,preprintnumbers,nopacs,floatfix,amsmath,amssymb]{revtex4}

\usepackage{graphicx}% Include figure files
\usepackage{dcolumn}% Align table columns on decimal point
\usepackage{bm}% bold math

\usepackage{amsmath}
\usepackage{graphicx}
\usepackage{amssymb}
\usepackage{mathrsfs}
\usepackage{bbm}
\usepackage{epsfig}
\newcommand{\be}{\begin{eqnarray}}
\newcommand{\ee}{\end{eqnarray}}

\begin{document}

%\pagenumbering{empty}
%\begin{titlepage}
%
\title{ Protein Regge Trajectories, Phase Coexistence and Physics  of Alzheimer's Disease }

%\vskip 5.0cm
\author{Andrey Krokhotin}
\email{Andrei.Krokhotine@cern.ch}
\affiliation{Department of Physics and Astronomy, Uppsala University,
P.O. Box 803, S-75108, Uppsala, Sweden}
\author{Antti J. Niemi}
\email{Antti.Niemi@physics.uu.se}
\affiliation{
Laboratoire de Mathematiques et Physique Theorique
CNRS UMR 6083, F\'ed\'eration Denis Poisson, Universit\'e de Tours,
Parc de Grandmont, F37200, Tours, France}
\affiliation{Department of Physics and Astronomy, Uppsala University,
P.O. Box 803, S-75108, Uppsala, Sweden}

\begin{abstract}
\noindent
Alzheimer's disease  causes severe neurodegeneration in the brain that  leads to a certain death. The defining  factor is the formation of extracellular senile amyloid plaques in the brain. However, 
therapeutic approaches to remove them have not been effective in humans, and so our understanding
of  the cause of Alzheimer's disease
remains incomplete. Here we investigate  physical
processes that might relate to its onset.  Instead of the extracellular amyloid, we  scrutinize the intracellular domain of its  precursor protein. We argue for a phenomenon  that has never before been discussed in the context of polymer physics: Like ice and water together,  the intracellular domain of the amyloid precursor protein forms a state  of phase  coexistence with another protein. This leads to an inherent instability 
that could well be among the missing pieces in the puzzle of Alzheimer's disease.
\end{abstract}


%\date{\today}

\maketitle

The neurological origin of Alzheimer's disease involves both genetic 
and environmental 
factors \cite{gene1}-\cite{gene5}. Its hallmark is the  accumulation of 
the  amyloid isoform A$\beta$42 into senile  plaques
 \cite{abeta1}-\cite{abeta2}. This
has motivated  several  immunotherapic  approaches to either 
clear or prevent the cerebral A$\beta$  deposits \cite{immu1}-\cite{immu3}. 
Unfortunately there are serious side-effects such as  the development 
of  aseptic meningoencephalitis \cite{meni1}-\cite{meni3}.  As a consequence we do 
not know whether targeting of  A$\beta$42 will cure or 
even  curb the disease in humans. The excess production of A$\beta$42 
might just be an indication that something else has gone wrong \cite{greek},  \cite{critic1}.

The A$\beta$42 is  a derivative of the transmembrane amyloid precursor 
protein (APP) by proteolytic cleavages \cite{abeta1}, \cite{swede}, \cite{appbeyond}.
APP comes in several isoforms, it is naturally  present in many  organs.
Its physiological function remains under a debate and the understanding of its proteolytic processing 
is also incomplete \cite{greek}, \cite{swede}, \cite{appbeyond}.  Both the dominant, non-amyloidogenic pathway 
and the disease related,  A$\beta$ generating amyloidogenic pathway produce isoforms of 
the APP intracellular domain (AICD)  \cite{swede},  \cite{cleave}.  We have scrutinized 
the physical properties of various  AICD  complexes, searching  for an intracellular agent that 
might  correlate with the onset of  the  anomalous A$\beta$42 production.  We 
identify an inherently unstable 
physical phenomenon  that  has never before been discussed in the context of polymer 
research. 

After the $\gamma/\epsilon$ cleavage of APP the AICD 
may  form a transcriptionally active state with the Fe65 family of 
nuclear multidomain adaptor proteins \cite{swede}, \cite{FEreview}-\cite{adaptor}.   
Even though  the  relation  between 
AICD and Alzheimer's disease is not yet understood, we know that 
AICD is a product and Fe65  is a participant in the proteolytic cleavage 
processing of APP into A$\beta$42. Not surprisingly  Fe65 already appears among the  potential 
therapeutic targets \cite{FEreview},  \cite{AICD}, \cite{immu3}. 

In isolation AICD is presumed to be an intrinsically unstructured protein \cite{unst}. 
However, upon binding to Fe65,  AICD can  assume a regular form that can be
analyzed with x-ray crystallography. Unfortunately,  the high precision 
data remains limited. Here we 
shall investigate the structure  with PDB  code 
3DXC (chain B)   \cite{exp}. It describes a complex of a 28 residue segment of AICD with the larger, 65 residue
host  Fe65. 
There  are also the closely related 3DXD and 3DXE,  these
can be analyzed similarly and with identical conclusions. We find that  
the complex appears to have physical properties that seems to set 
it apart from all but a very few oligomers.  It  is  an example of an apparently previously unrecorded
but seemingly systematic phenomenon of {\it protein (polymer) phase coexistence}:  Like ice
with water  the  two proteins are in two different  phases.  
As such,  an oligomer that displays the rare and inherently unstable phenomenon of phase coexistence  
is for sure an interesting object for future research. But the delicate balance of the AICD/Fe65 complex  
has the supplementary  potential of  being an important 
piece in the puzzle to find a cure for Alzheimer's disease.

%\section{Method}

The phases of a protein and more generally those  of a polymer, are characterized by  
their fractal (Hausdorff) dimension. This is an order parameter that can be  computed by inspecting the 
scaling properties of the radius of gyration $R_g$.  Asymptotically, in the limit where the 
number $N$ of monomers 
becomes very large \cite{degennes}
\begin{equation}
R_g \ = \ \frac{1}{N} \sqrt{ \frac{1}{2} \sum_{i,j} ( {\bf r}_i 
- {\bf r}_j )^2 }  \ \approx  \ R_0\cdot N^{\nu} 
%\ = \ 3.25 \cdot N^{0.35} \ \ \ (\.A^2)
\label{R}
\end{equation} 
Here $\mathbf r_i$ are the coordinates of the backbone $C_\alpha$,  the pre-factor $R_0$ is an 
effective inter-monomer distance that is independent of $N$, and $\nu$ is the  compactness index that
equals the inverse fractal dimension of the backbone. The remarkable property of (\ref{R}) is that $\nu$ is a {\it universal} quantity  \cite{degennes}, \cite{schafer}.  Different values of $\nu$ correspond 
to different phases,  and once we know $R_0$  we can  unanimously  compute the radius of gyration by simply 
counting the number of monomers. 
All the effects of temperature and  chemical microstructure and all atomary level details of a polymer are
contained in the value of the {\it a priori}  non-universal and in principle computable pre-factor  $R_0$. 

The relation (\ref{R}) becomes truly precious {\it only} in those exceptional circumstances 
where $R_0$ assumes no more than a small number of different values.  We argue that this is indeed
the case in proteins:  When $N$ increases,  proteins
become increasingly uniform in their chemical composition. Consequently 
it makes sense to employ (\ref{R}) to study their phase structure.
Different, clearly identifiable trajectories  (\ref{R}) that are labelled by the 
different well defined values of $R_0$  are then the protein  analogs of  the Regge trajectories  in high energy physics \cite{regge}. 

Proteins and other polymers \cite{degennes} \cite{schafer} have four major phases: 
Under physiological conditions  and in other bad solvents a protein collapses into a space filling 
conformation with  $\nu \approx 1/3$. For a fully flexible chain we have  the $\Theta$-point value $\nu \approx 1/2$ while in the  self-avoiding 
random walk phase we have the Flory value $\nu \approx 3/5$.  
Finally,  when $\nu \approx 1$ the protein 
looses its inherently fractal structure and becomes
like a one dimensional rigid rod. Examples of this phase are monotonous $\alpha$-helices and $\beta$-strands that
have no additional twists, turns or loops.

\begin{figure}%[!hbtp]
  \begin{center}
    \resizebox{15.cm}{!}{\includegraphics[]{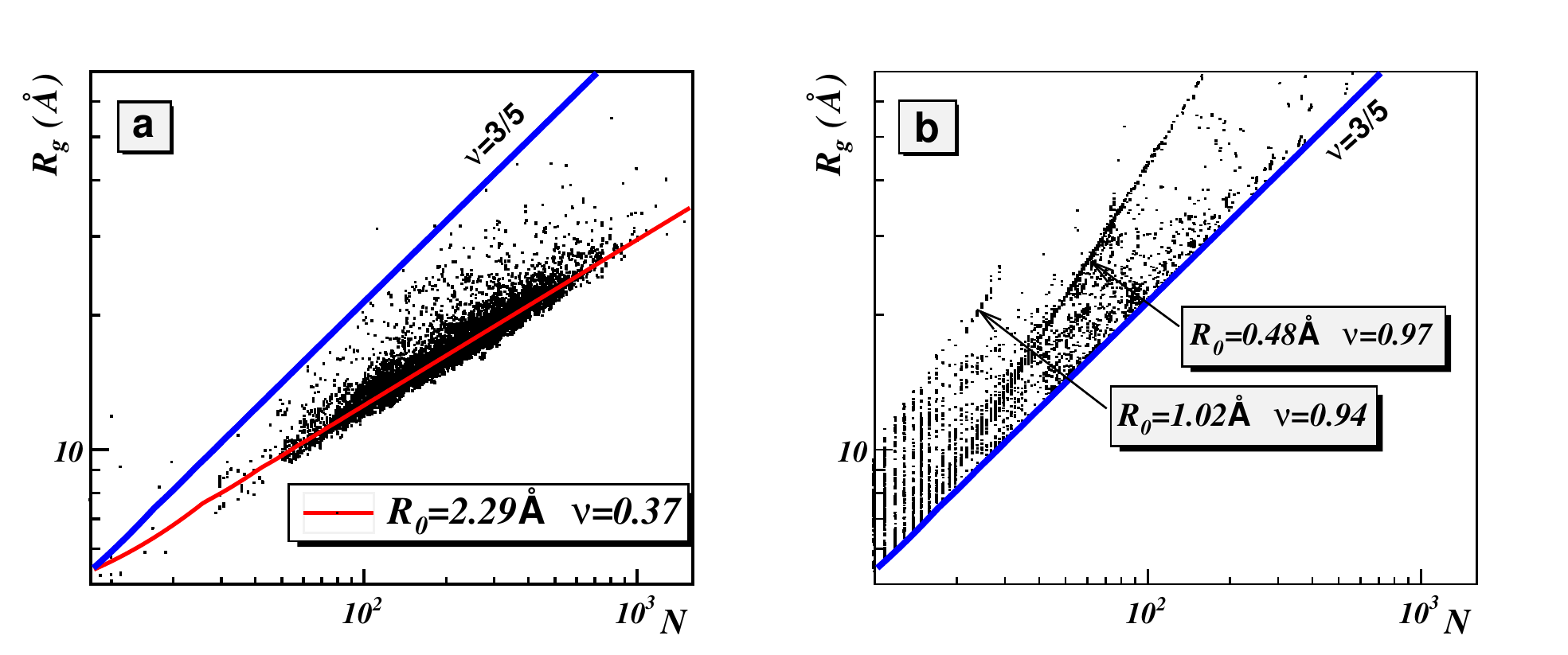}}
    \caption{ a) The ($N, R_g$) distribution of all single chain PDB proteins with resolution less than 2.0 \.A 
    and with less than 30$\%$ homology equivalence. The lower line is the Regge
     trajectory for mostly $\alpha$-helical proteins and  the 
     top line is a  $\nu = 3/5$ Flory line; there are practically no single chain 
     entries above this  line.
    b) All proteins currently in PDB that are above the Flory line. The two Regge trajectories (3) are clearly visible.   }
    \label{fig1}
  \end{center}
\end{figure}
In Figure 1a we plot all individual single chain 
proteins in PDB  that have resolution less than 2.0 \.A and homology equivalence which is less than 30$\%$.
With a few exceptions they 
assemble around a Regge  trajectory with $\nu \approx 1/3$.  We also plot
the dominant mostly-$\alpha$-helical trajectory with $R_0 \approx  2.29$ and   $\nu \approx 0.37$; 
The numerical values are slightly different for the different protein 
subclasses like  mostly-$\alpha$-helical,  mostly-$\beta$-sheets  {\it etc.}  and this  fine 
structure has been discussed in \cite{add1}, \cite{add2}. 

When we extend our analysis to individual chains within oligomers
we find  two previously unobserved clearly visible Regge trajectories (Figure 1b).
These two  trajectories  are
\begin{equation}
R^{(2)}_g \ \approx \  \  0.48 \cdot N^{0.973}  \ \ \ \ \ \& \ \ \ \ \ 
R^{(3)}_g \ \approx \  \ 1.02 \cdot N^{0.94}
\label{R34}
\end{equation} 
These trajectories both have  $\nu$ very close to one. Thus  
they must be in the same universality  class, and the difference is a  finite size effect. 
This is the universality class   of one dimensional rods and sticks. Unlike the  other three
polymer phases, it has  no fractal structure. 

The trajectory $R^{(2)}_g$   includes several
membrane proteins and viral capsomers, an example of the latter is 1AIK in PDB. The trajectory 
$R^{(3)}_g$ is mainly populated by collagen proteins such as for example 2CUO in PDB.  
In both trajectories the oligomers commonly consist of several individual chains 
that are each located on the same Regge trajectory.   Their mutual interactions 
provides  a supportive lattice structure that protects
the individual chains against a  collapse into the  $\nu \approx  1/3$ phase. 

Remarkably, there are also protein complexes in the Regge trajectories of Figure 1b
that do not follow the structural pattern of collagens, membrane proteins or viral capsomers. In particular,
we have found that there is a small number of  oligomers  that are composed  of proteins on  {\it different} Regge 
trajectories. These complexes consist of (host) sub-chains that are in a
$\Theta$-point  trajectory $\nu \approx 1/2$  and  (guest)  sub-chains on a similarly
uncollapsed  $\nu \approx 1$ trajectory of Figure 1b. 
These oligomers are examples of a previously unrecorded  physical phenomenon of {\it protein 
phase coexistence}: The interaction between the  host and the guest  
provides a support that maintains each of them  in an 
inherently unstable uncollapsed conformation. 

We have found two different classes of such phase coexistent  oligomers in PDB. The first class consists of an apparently single protein but 
with multiple sub-chains that are in different phases. The present PDB codes for these proteins  
are  1WDC, 1G72, 1GOT, 1HTR, 1LTS, 2FP7, 2RIV,  3ABK,  3ARC, 3CX5, 3DBO. Here we concentrate on 
the  second class,  formed by complexes with  two or more {\it a priori} different proteins. 
The present PDB codes are 1L2W,  1JDH, 1TH1, 2F8X,  2EPV, 2PRR, 2BFX, 2D7C, 2VGO, 2K8F, 2QKH,  3EGG,  3HTU, 3HPW, 3IXS and 3DXC (3DXD, 3DXE). 
In Figure 2 we display the distribution of the individual sub-chains of the second class in the ($N,R_g$) plane, they clearly gather around a $\nu \approx 1/2$ 
Regge trajectory and the $R^{(2)}_g$ trajectory in (\ref{R34}).
\begin{figure}%[!hbtp]
  \begin{center}
    \resizebox{15.0cm}{!}{\includegraphics[]{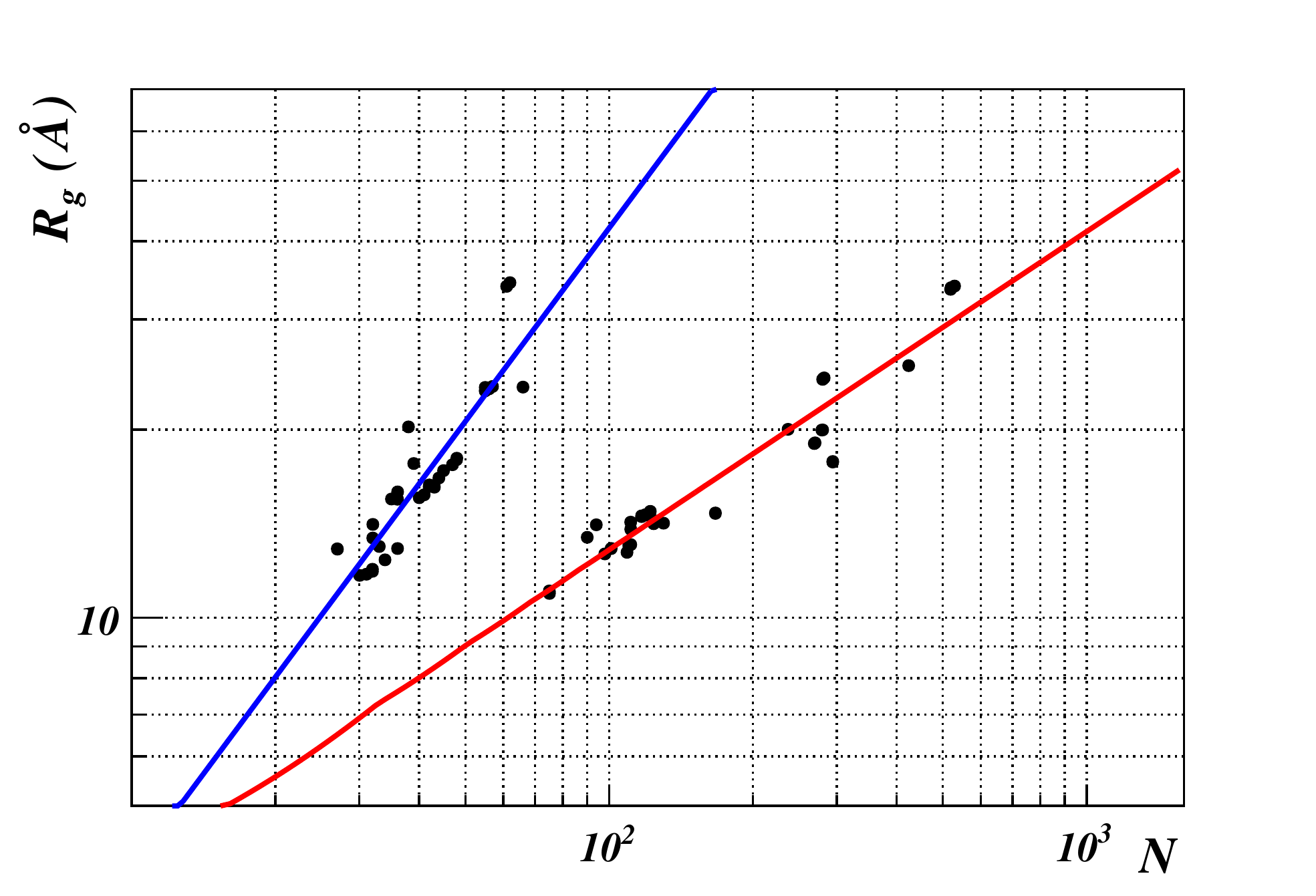}}
    \caption{The distribution of individual chains on the ($N,R_g$) plane  in our second class of phase coexistent  complexes. 
    The data clearly accumulates around the top line that describes the Regge trajectory  $R^{(2)}_g$ and the 
    bottom line that describes a $\Theta$-point Regge trajectory with  
    best fit  values $R_0 = 1.234$
    and $\nu = 0.508$.
     }
    \label{fig2}
  \end{center}
\end{figure}
Biologically, the two most  notable  are 
2K8F and 3DXC (3DXD, 3DXE). The former is a bound state  of  the 
"molecular interpreter" p300 \cite{p300}, \cite{unst}  in the Regge trajectory $R^{(2)}_g$,  with  the tumor suppressing protein p53 in the $\nu \approx 1/2$ trajectory. The second is the one of interest here, the Alzheimer related AICD/Fe65  complex with AICD in the trajectory  $R^{(2)}_g$  and Fe65 in the $\nu \approx 1/2$ trajectory.  
We now proceed to analyze the peculiar physical properties of  the Alzheimer related 
AICD of the second  complex. 

%\section{results}

In Figure 3a we display the C$_\alpha$ backbone Frenet frame  bond and torsion angles of the AICD protein in  3DXC.
\begin{figure}%[!hbtp]
  \begin{center}
    \resizebox{15.0cm}{!}{\includegraphics[]{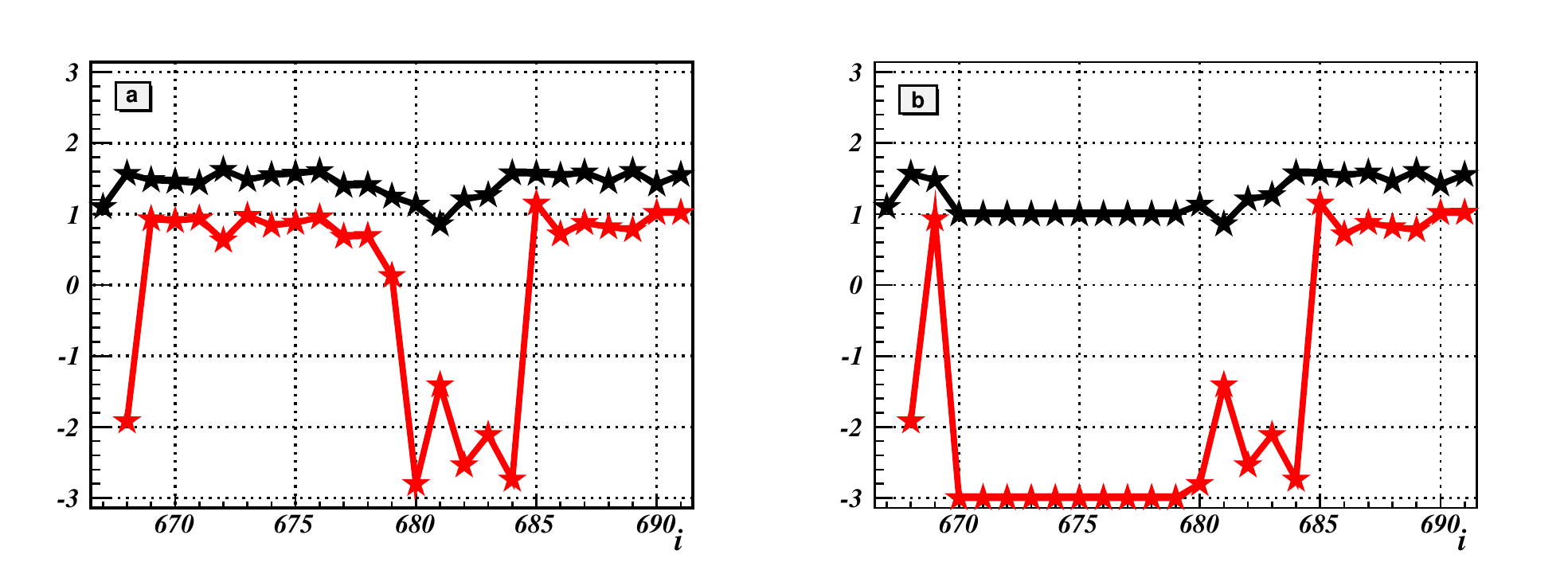}}
    \caption {a) The spectrum of backbone Frenet frame bond angles $\psi_i$ (top line) and 
    torsion angles $\theta_i$ (bottom line) for the AICD component of 3DXC (chain B).  b) The same 
    spectrum after we have translated the first loop  so that it becomes locked  by the proline at site 669, 
    as described in the text.  We use PDB indexing for the sites. }
    \label{fig3}
  \end{center}
\end{figure}
This Figure  reveals that the AICD  consists of two very closely located loops that are separated  from each other by a  
very short $\beta$-strand. We can describe the profile of each of these angles using the  soliton solution of 
nonlinear Schr\"odinger equation, for the backbone bond 
angles we have \cite{maxim}, \cite{peng}
\begin{equation}
\psi_i  
% \arccos
%\left \{ 
%\frac{  
%{\bf r}_{i+1} - {\bf r}_{i } } {  | {\bf r}_{i+1} - {\bf r}_i |  }   \cdot 
%\frac{  {\bf r}_{i} - {\bf r}_{i-1 } } {  | {\bf r}_{i} - {\bf r}_{i-1} | } \right \}
= \frac{ m_{1}   \cdot e^{ c_{1} ( i-s)  } - m_{2} \cdot e^{ - c_{2} ( i-s)}  }
{e^{ c_{1} ( i-s) } +  e^{ - c_{2} ( i-s)}  }  \  \in [-\pi, \pi] \  \mod(2\pi)
\label{bond}
\end{equation}
while the backbone torsion angles are computed in terms of the bond angles from 
\begin{equation}
\theta_{i} =    - \frac{1}{2} \frac{ a }{1+b \cdot \psi_i^2}   \in [-\pi, \pi] \  \mod(2\pi) 
\label{tors}
\end{equation}
and the parameter values are listed in Table I: The corresponding ($\psi_i, \theta_i$)  profiles (\ref{bond}), (\ref{tors})
describe the first loop at sites 676-683 (we use PDB indexing) with  RMSD accuracy of 0.29 \.A and the second loop
at sites 681-688 with RMSD accuracy of 0.17 \.A. Both accuracies are substantially better than the experimental
B-factor accuracies.
\begin{table}[tbh]
\begin{center}
\caption{Parameter values for the two loops in Figure 2. }
\vspace{3mm}
\begin{tabular}{|c|cccccc|}
\hline
loop &  $m_1$ & $m_2$ & $c_1$ & $c_2$ & $s$  & match\\
\hline
676-683 & 51.517 & 51.766  & 2.984 & 2.983 & 679.909 & 177  \\
\hline
681-688  & 39.274 & 38.617 & 3.327  & 3.347 & 682.174 & 896 \\
\hline
\end{tabular}
\end{center}
\label{solenoid}
\end{table}
In Table I we also list the number of times each of these loops appear in PDB with RMSD accuracy  0.5 \.A or better. 
Both are {\it abundant} in the $\nu \approx 1/3$ Regge trajectory of collapsed proteins. 
 In fact, at the outset there is nothing in the secondary structures of this AICD fold that  appears unusual for a protein in the collapsed $\nu \approx 1/3$ phase. Nevertheless it is  very accurately, almost exactly, 
located on the $\nu \approx 1$ Regge
trajectory  $R^{(2)}_g$.  

Since the compactness index $\nu$ is universal and can only have definite discrete values,  any {\it continuous and local} deformation of the protein shape can 
never cause any kind of  {\it discontinuous} transition such as a jump between the two phases $\nu \approx 1$ and $\nu \approx 1/3$.  This makes 
the  present combination of the two loops in AICD highly unusual. Even if we continuously translate the two loops apart from each  other along the backbone by shifting the value $s$ in (\ref{bond}) that determines the position of the center of the loop,
we can never  reach a collapsed  Regge trajectory but will always remain in the $\nu \approx 1$ phase.

A scrutiny of the amino acid structure reveals that AICD has a proline at site 669.
Since proline often acts as an anchor of a loop in a protein in isolation, we propose that the presence of Fe65 prevents the first loop from sliding towards  its natural position, where it becomes attached with Pro(669). Note that there is another proline at the site 685 that appears  to stabilize the position of the second loop. Using the explicit profile (\ref{bond}),
(\ref{tors})  we  investigate what might happen if the 
first soliton starts sliding  towards Pro(669) along the backbone. 
For this we shift the value of the parameter $s$ accordingly. 
In Figure 4 we show how the radius of gyration $R_g$  of the AICD
depends on the position of the first loop as we slide it  towards Pro(669) while keeping 
the second loop  anchored by Pro(685); the 
final ($\psi_i, \theta_i$) profile is displayed in Figure 3b.
\begin{figure}%[!hbtp]
  \begin{center}
    \resizebox{15.0cm}{!}{\includegraphics[]{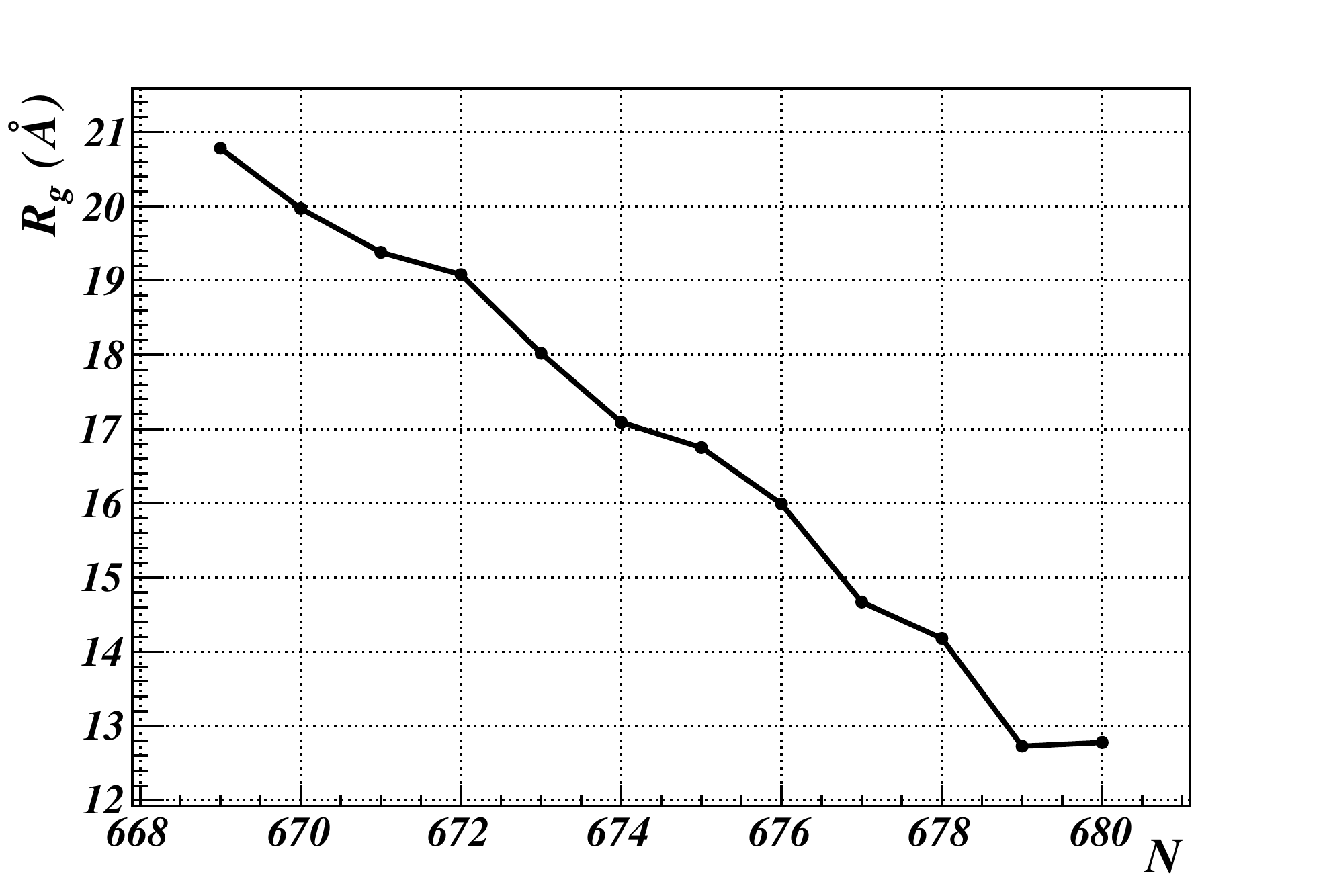}}
    \caption{The evolution of the radius of gyration for the AICD in 3DXC (chain B), during the translation of its first loop to the proline at site 669; 
    see also Figure 3b. }
    \label{fig4}
  \end{center}
\end{figure}
We find that  $R_g$ increases monotonically when the two loops drift apart. When the first loop reaches the position where 
it  becomes locked by Pro(669),  the ensuing AICD configuration has relocated itself from the
Regge trajectory $R^{(2)}_g$ to the Regge trajectory $R^{(3)}_g$.  Since it should be {\it highly} natural for a protein to always try and locate itself on a Regge trajectory, 
we propose that these are the two likely configurations of AICD in the complex with Fe65.  The presence of two natural alternatives
is an indicative of a genetic switching mechanism. It would be highly  interesting to find out  how the biological function of the 
AICD/Fe65 complex differs  between these two 
unfolded  conformations  of AICD, when the complex becomes translocated
to the nucleus and  participates in gene  transcription.  Is there a correlation with the onset of  Alzheimer's disease?
Moreover, the genetic switch could even operate solely around Pro(669), the first soliton could conceivably be on either side of this proline.
Suppose it  is located  on the other side of Pro(669) when the $\gamma/\epsilon$ cleavage takes place. It could then
become part of A$\beta$. Could this cause the formation of  senile amyloid  plaques?

In isolation, the $\nu \approx 1$  phase of AICD  must be extremely unstable under {\it in vivo} conditions, we have not found {\it any }
single strand protein above our  $\nu = 3/5$ Flory line. Since its two solitons are very common among $\nu \approx 1/3$ proteins, we predict that an isolated  AICD becomes subject to a phase transition 
that takes it into the collapsed  $\nu \approx 1/3$  trajectory.  
Thus we  propose that when the two proteins are disengaged,  AICD collapses either in a process where the two loops first 
pair-annihilate each other and a new loop  structure is formed to bring about the phase transition, or alternatively there could be  the formation of a new loop near Pro(669). Alternatively,  AICD enters
a highly unstructured and dynamic state where the first loop  bounces back and forth between the two prolines,
causing AICD to oscillate between the two $\nu \approx 1$ Regge trajectories.  
Other alternatives also exist. For example the first loop could become locked by Pro(669),
and the relatively long $\beta$-strand could then buckle to form a new loop. We propose xperiments 
are designed to find out the properties of AICD under various bad solvent conditions.

We conclude that some of the 
proteins that are involved in the onset of Alzheimer's disease can be set apart by their  rare
physical properties.   In particular the presence of a protein oligomer, and more generally a polymer complex,
with a phase co-existence should be a challenge for future investigations.
In particular,  if it turns out that the origin of Alzheimer's disease is due to the  ensuing  instabilities. 
%{\bf References}
%\vfill\eject


\begin{thebibliography}{99}

\bibitem{gene1}  M.C. Chartier-Harlin {\it  et.al.}  
%Early onset AlzheimerÕs disease caused by mutations at codon seven hundred
%and seventeen of the beta-amyloid precursor protein gene. 
Nature {\bf 353} (1991) 844
%844-846

\bibitem{gene2} R. Sherrington {\it et.al.}
%  Cloning of a gene bearing missense mutations in early-onset familial AlzheimerÕs disease.
Nature {\bf 375} (1995) 754
%, 754-760 

\bibitem{gene3}E.  Levy-Lahad {\it et.al.}
% Candidate gene for the chromosome 1 familial AlzheimerÕs disease locus. 
Science {\bf 269} (1995) 973
%, 973-977

\bibitem{gene4} M. Gatz {\it  et.al.}
%   Role of genes and environments for explaining Alzheimer disease. 
Arch. Gen. Psych. {\bf 63} (2006) 168
% 168-174 

\bibitem{gene5}  W.B. Grant, A. Campbell, R.F. Itzhaki and J. Savory, 
%The significance of environmental factors in the etiology of Alzheimer's disease.
Journ. Alz. Dis. {\bf  4} (2002) 179
%179-189 

 \bibitem{abeta1} G.G. Glenner, C.W. Wong, V. Quaranta and E.D. Eanes,
 % The amyloid deposits in AlzheimerÕs disease: their nature and pathogenesis. 
 Appl. Pathol. {\bf 2} (1984) 357
 % 357-369 

\bibitem{hardy}  J. Hardy and D.  Allsop,  
%Amyloid deposition as the central event in the aetiology of AlzheimerÕs disease. 
Trends Pharmacol. Sci. {\bf 12} (1991) 383
% 383-388

\bibitem{abeta2} J. Hardy and D.J. Selkoe, 
%The Amyloid Hypothesis of AlzheimerÕs Disease: Progress and Problems on the Road to Therapeutics.  
Science {\bf 297} (2002) 353
% 353-356 

\bibitem{immu1} H.J. Fu,  B. Liu,  J.L. Frost and C.A. Lemere, 
% Amyloid-$\beta$ Immunotherapy for Alzheimer's Disease. 
CNS Neurol. Disord. Drug Targets {\bf 9} (2010) 197
% 197-206

\bibitem{immu2} K. Rezai-Zadeh, D. Gate, G. Gowing and T. Town, 
%How to Get from Here to There: Macrophage Recruitment in Alzheimer's Disease. 
Curr. Alzheimer Res. {\bf 8} (2011) 156
% 156-163 

\bibitem{immu3} E.D. Roberson and L. Mucke, 
% 100 Years and Counting: Prospects for Defeating AlzheimerÕs Disease. 
Science {\bf 314} (2006) 781
% 781-784

 
 \bibitem{meni1} J.-M. Orgogozo {\it et.al.} 
%Subacute meningoencephalitis in a subset of patients with AD after A42
%immunization. 
Neurology {\bf 61} (2003)  47
% 47-54

\bibitem{meni2} C. Holmes {\it  et.al.} 
%Long-term effects of Abeta42 immunisation in AlzheimerÕs
%disease: follow-up of a randomised, placebo-controlled phase I trial.
Lancet {\bf 372} (2008) 216
% 216-223

\bibitem{meni3} N.F. Schor, 
% What the halted phase III $\gamma$-secretase inhibitor trial may (or may not) be telling us.
Ann. Neurol. {\bf 69} (2011) 237
% 237-239 

\bibitem{greek} N.K. Robakis, 
% Mechanisms of AD neurodegeneration may be independent of A$\beta$ and its derivatives. 
Neurobiol. Aging {\bf 32} (2011) 372
%  372-379 

\bibitem{critic1} A. Mudher and S. Lovestone,  
%AlzheimerÕs disease Ð do tauists and baptists finally shake hands? 
Trends Neurosci. {\bf 25} (2002) 22 
% 22-26


\bibitem{swede} K.T. Jacobsen and  $\AE$.K.  Iverfeldt,  
 %Amyloid precursor protein and its homologues: a family of proteolysis-dependent receptors. 
 Cell. Mol. Life Sci. {\bf 66} (2009) 2299
 % 2299-2318 
 
 
 \bibitem{appbeyond}  H. Zheng and E.H.  Koo,
% The amyloid precursor protein: beyond amyloid. 
Molec. Neurodeg. (online)  {\bf 1} (2006) 1
% 1-12


\bibitem{cleave} S.B. Roberts, J.A. Ripellino, K.M. Ingalls, N.K. Robakis and K.M. Felsenstein,
%Non-amyloidogenic cleavage of the $\beta$-amyloid
%precursor protein by an integral membrane metalloendopeptidase.
J.  Biol. Chem. {\bf 269} (1994) 3111
% 3111-3116


\bibitem{FEreview} D.M. McLoughlin, C.J. Christopher and C.C.J. Miller, 
% The FE65 Proteins and AlzheimerÕs Disease. 
Journ. Neurosci. Res. {\bf 86} (2008) 744
% 744-754

\bibitem{AICD}  T. M\"ullera, H.E. Meyera, R. Egenspergerb and K. Marcusa,   
%The amyloid precursor protein
 % intracellular domain (AICD) as modulator of gene expression, apoptosis, and cytoskeletal dynamics - Relevance for % Alzheimer's disease.
 Prog.  Neurobiol. {\bf 85} (2008) 393
 %  393-406 
  
 \bibitem{adaptor} S.L. Sabo {\it et.al.}  
 %Regulation of b-amyloid
%secretion by FE65, an amyloid protein precursor-binding protein.
J. Biol. Chem. {\bf 274} 91999) 7952
% 7952-7957 
 
 \bibitem{unst}  H.J. Dyson and P.E. Wright, 
%Intrinsically unstructured proteins and their functions. 
Nature Rev. Molec. Cell Biol.  {\bf 6} (2005) 197
%  197-208  

\bibitem{exp} J. Radzimanowski {\it et.al.}  
%Structure of the intracellular domain of the amyloid precursor protein in complex with Fe65-PTB2.
Embo Rep. {\bf 9} (2008) 1134
%1134-1140 

\bibitem{degennes} P.G. De Gennes,  {\it Scaling Concepts in Polymer Physics} (Cornell  University Press, Ithaca, 1979)

\bibitem{schafer} L. Sch\"afer, {\it Excluded Volume Effects in Polymer Solutions, 
as Explained by the Renormalization Group} (Springer Verlag, Berlin, 1999)


\bibitem{regge} P.D.B. Collins, {\it An 
Introduction to Regge Theory and High-Energy Physics.}  (Cambridge University Press, Cambridge, 1977)


\bibitem{add1} T.G. Dewey, 
%Protein structure and polymer collapse. 
Journ.  Chem. Phys. {\bf 98} (1993) 2250
% 2250-2256 

\bibitem{add2} L. Hong and J.  Lei, 
%Scaling Law for the Radius of Gyration of Proteins and Its Dependence on Hydrophobicity. 
Polym. Sci. {\bf B47} (2009) 207
% 207-214



\bibitem{p300} J.L. Smith {\it et.al.} 
%Kinetic profiles of p300 occupancy in vivo predict common features of promoter structure and
%coactivator recruitment. 
Proc. Natl Acad. Sci. USA {\bf 101} (2004) 11554
% 11554-11559 

\bibitem{maxim}  M. Chernodub, S. Hu and A.J.   Niemi, 
%Topological Solitons and Folded Proteins.
Phys. Rev. {\bf E82} (2010) 011916
% 011916-011920 

\bibitem{peng} S. Hu, A.  Krokhotin, A.J. Niemi and X.  Peng,   
% Towards quantitative classification of folded proteins in terms of elementary functions.
Phys. Rev. {\bf E83} (2011) 041907
% 041907-041912 



\end{thebibliography}
\end{document}